# Magnetic behavior of the metal organic framework solid: $[(CH_3)_2NH_2][Co(HCOO)_3]$

K. Vinod[a], C.S. Deepak[b], Shilpam Sharma[a], D. Sornadurai[b], C. S.Sundar[a] and A. Bharathi[a]

In this study we examine the phase transitions in single crystals of $[(CH_3)_2NH_2]Co(HCOO)_3]$, using magnetization and specific heat measurements as a function of temperature and magnetic field. Magnetisation measurements indicate a transition at 15 K that is associated with an antiferromagnetic transition. The results of the isothermal magnetization versus  magnetic field curves demonstrate the presence of a single-ion magnet phase, coexisting with antiferromagnetism. A peak in specific heat is seen at 15 K, corresponding to the magnetic transition. The enthalpy of the transition evaluated from the area under the specific heat peak decreases with the application of magnetic field of upto8 T. This is suggestive of long range antiferromagnetic magnetic order, giving way to single-ion magnetic behavior under external field. In experiments at high temperatures, corresponding to the well-known structural transition in this system, the specific heat measurements, shows a peak at ~155K, that is insensitive to applied magnetic field. The magnetisation in this temperature range, while it exhibits a paramagnetic behavior, shows a distinct jump that has been attributed to a spin-state transition of $Co^{2+}$ associated with the structural transition.

## Introduction

The interests in the metal organic framework (MOF) solids, the organic-inorganic hybrids stems from their enormous  potential for applications in gas storage, catalysis, drug delivery, sensors, electronic devices, light emitting devices, and as anode in Li-based batteries [1-5]. The family of MOF solids having a perovskite architecture, Di-Methylamine Metal Formamide, $[(CH_3)_2NH_2][M(HCOO)_3]$ with M = Mn, Fe, Co and Ni, has attracted considerable attention due to the coupling between  their magnetic, dielectric and structural properties [6-10]. Wang et.al [6], in their first study, have shown that the system exhibits long range canted antiferromagnetic order at low temperatures (Indicate the transition temperature, if possle). Jain et al [7] have shown that there have shown that the dielectric transition from paraelectric to an antiferroelectric phase, occurring at ~ 160 K  that is coincident with the structural transition from rhombohedral to monoclinic. The later has been attributed to the orientational ordering of N atoms (See Fig.1.). Specific heat measurements [7] indicate anomalies at 160 K and 8 K corresponding to the electrical and magnetic transitions.

In a recent interesting study on DMA– Fe – formamide single crystals, Tian et al [11] have shown the occurrence of a canted antiferromagnetic (AFM) transition at 20 K followed by a drop in magnetization at 8 K, attributed to a single ion magnet, showing resonant quantum tunnelling,. The presence of two co-existing magnetic phases has been rationalized in terms of the structure (see Fig.1) that leads to two competing magnetic interactions, one through the pure formate linker, and the other due to the formate coupled with the DMA⁺cation via a hydrogen bond (N-H…O).  Dielectric constant measurements in this system [12,13] show an increase below the magnetic transition at 20 K, pointing to a magneto-electric coupling. The dielectric anomaly at the structural transition temperature of ~ 160 K is also affected by the application of external magnetic field. These results imply the cross coupling between electric and magnetic orders in this mutiferroic metal-organic framework system.

The aim of our study is to investigate phase transitions corresponding to magnetic order and order-disorder structural transition in the corresponding Co-formamide system, viz., $(CH3)2NH2Co(HCOO)3$. These investigations were carried out to see if the Co formamide system bears any similarity to the Fe compound and also investigate if $Co^{+2}$ ,that has a propensity to exist in various spin states [14] has any influence on the observed phase transitions in the material. Towards this, we first synthesize single crystals of $[(CH_3)_2NH_2]Co(HCOO)_3]$ by the already established route [7]. We investigate the magnetic property of the tiny single crystal using a SQUID vibrating sample magnetometer, in external magnetic fields of upto 4 Tesla. The magnetization data, are followed up with highly sensitive, specific-heat measurements, carried out on a 250 micron sized single crystal, by ac calorimetry, in the temperature range of 4 to 300 K, and in magnetic fields upto 8 T.  These results point to the presence of two coexisting magnetic phases, as in the Fe system [11]. Further, the magnetization measurements across the order-disorder structural transition, shows a distinct jump, not reported in other systems,  that has been attributed to spin state transition in $Co^{2+}$.

## Experimental

The synthesis of Co compound was carried out using a solvothermal approach [7], in which 30 ml of di-methyl formamide and 30 ml of millipore water are taken in a 100 ml beaker and 5 mmol of cobalt chloride, hexahydrate ($CoCl_2.6H_2O$) was added to this solution. The crystals readily dissolve in this solution to give a deep pinkish red color. This solution was filled into Teflon lined autoclave and heated treated at 140 $^0C$ for 3 days. The furnace was switched off and the autoclave was allowed to remain in the furnace. The autoclave was removed from the furnace when it was at 42 $^oC$ and allowed to cool in air. After the autoclave was cooled to room temperature, it was opened and the supernatant was poured into another 100 ml beaker and left for three days, at ambient temperature. Crystals formed were primarily attached to the walls of the beaker. Crystals were filtered and washed with alcohol. Shiny crystals of flat and cuboidal morphologies were seen to form, as discerned under the microscope. From powder XRD data, we could infer that the crystals with flat morphology were $Co(HCOO)_2.2H_2O$ [15] and those with cuboidal morphology were $[(CH_3)_2NH_2]Co(HCOO)_3]$. The crystals with cuboidal morphology (see inset Fig.1) were separated and used in further experiments. Powder XRD measurements at room temperature, was carried out in the BL-12 beam line of the Indus-2 synchrotron in Indore India, using 16 keV x-rays [16]. The powdered samples were placed in a circular depression made on Kapton tape. The data were collected on a MAR3450 image plate detector in the transmission geometry. Fit2D program [17] was employed to convert the 2D image from the image plate detector to 1D Intensity versus 2θ plot. The experimental XRD pattern obtained is shown in the main panel of Fig.1. The diffraction data was analysed using trigonal structure of space group (R -3 c (167)) and the lattice parameters, extracted employing the GSAS program[18,19] are a =b = 8.2012$\pm$ 0.0007 A and c= 22.547$\pm$0.003 A. These values lie between those obtained for Fe and Zn formamide compounds [20], as would be expected.

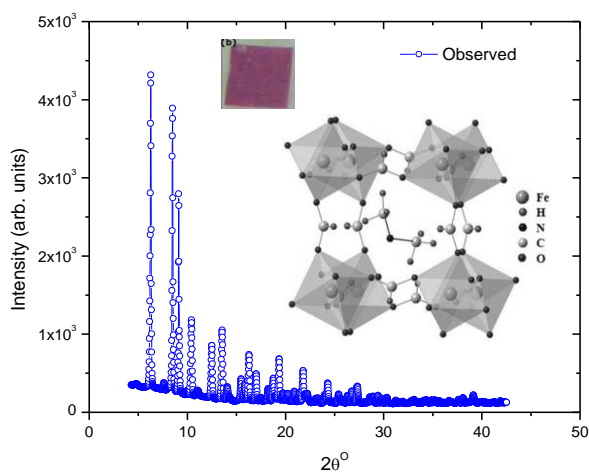

Fig.1 XRD pattern of the ground crystals $[(CH_3)_2NH_2]Co(HCOO)_3]$. Inset shows a schematic of the structure taken from Ref. [11]. The metal atoms located at the corners of the cube and are connected via the oxygen atoms of the formate linkers. The Di Methyl Amine cation is located at the centre of the cavity and is hydrogen-bonded to the formate frame via its amino H atoms to the formate O atoms. The optical microscope image of the crystal of lateral dimension ~0.2 mm used for the physical property measurements.

The magnetization measurements were carried out in a Quantum Design Ever-Cool SQUID vibrating sample magnetometer capable of a temperature variation of 1.6 K to 320 K and field variation of 0 to 7 Tesla, with a base sensitivity of $10^{-7}$ emu. The magnetization measurements were carried out in the zero field cooled (ZFC) and field cooled (FC) protocols. For the temperature scan during the ZFC and FC measurements, the temperature was varied at the rate of 2 K/min upto 50 K and at 8 K/min from 50 K up to 180 K. For the scans as a function of magnetic field the ramp rate of magnetic field was done at the rate of ~50 Oe/ s. The measurement of specific heat on the single crystal is carried out by ac calorimetry, in the 4 K to 300 K temperature range, in external fields upto 8 Tesla, using a Cryogenic-make insert, to the existing vibrating sample magnetometer. In this ac calorimetric technique a sinusoidal current (11 Hz) is passed through a heater and the temperature rise is monitored by thermopiles phase sensitively [21]. The heater and the thermopiles are a part of a custom made chip and are in close proximity to the 250 micron sized crystal. The heat capacity using this system has a sensitivity of ~nJ/K, enabling measurement on very small crystals. The ramping of temperature for the heat capacity versus temperature measurements was done at a rate of 0.5 K/min.

## Results and discussions

For the magnetization measurements, a single cuboidal crystal of ~0.2x0.2x0.16 mm was placed on the puck of the sample holder in which the face of the crystal (012), which happens to be the growth direction, was perpendicular to the applied field direction. The magnetization versus temperature data with a measuring field of 0.05 T measured under Zero Field Cooled (ZFC), field cooled (FC) protocols are shown in Fig.2. It is evident from the figure that there is clear bifurcation of the ZFC and FC curves indicative of a phase transition at 15 K. As seen from the figure, the difference between the ZFC and FC curves diminish with increase in field and in particular the step like behavior seen at ~5 K gives way to a gradual increase in magnetization with decrease in temperature. The 15 K magnetization anomaly, however, becomes a broad cusp, under large measuring magnetic field of 4T, implying that the transition is antiferromagnetic in nature, similar to that observed in Fe compound [11]. In addition, the low temperature step like increase at ~ 8 K in magnetization seen the FC data at upto 0.1 T, gradually shows a monotonic increase with decrease in temperature. The results of isothermal magnetization as a function of magnetic field are displayed in various panels in Fig.3. In the low field region, a jump in magnetization, is seen, whereas, at higher fields the M versus H is linear, characteristic of the antiferromagnetic nature of the magnetic ground state [11]. As seen in the measurements in the Fe compound [11], the low field jump in magnetization can arise due to free spins aligning easily with an applied magnetic field. Thus from our measurements ~~is~~it can be inferred that the presence of two coexisting magnetic phases could be a generic feature in all the transition metal compounds belonging to the class $[(CH_3)_2NH_2][M(HCOO)_3]$ where M is a magnetic transition metal ion. However, it should be noted the double hysteresis loops seen in the Co compound is not as pronounced as that seen in the corresponding Fe compound [11], which may be due to the difference in the strength of the magnetic coupling between the – magnetic species in the two compounds.

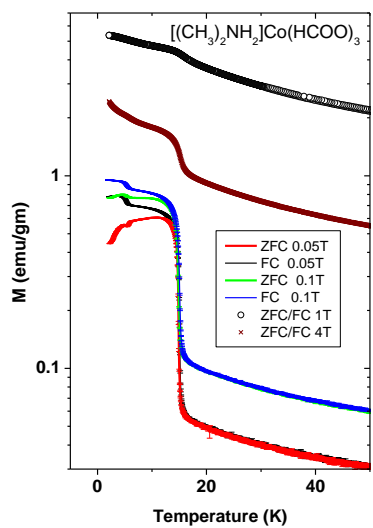

Fig.2 The magnetization versus temperature in the 1.6 to 50 K temperature range, with measuring fields indicated, ZFC and FC curves are separated for measuring fields of 0.1 T and 0.05 T.

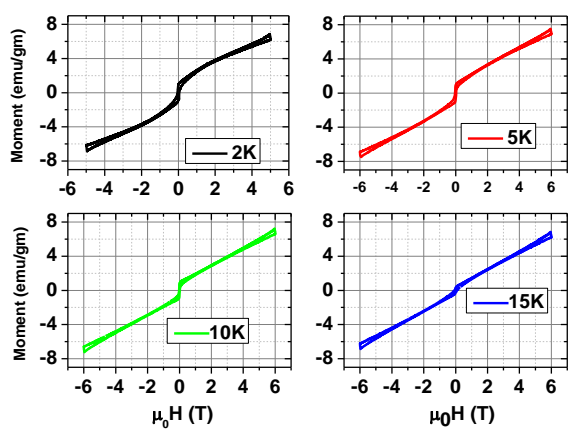

Fig.3 Isothermal Magnetisation versus magnetic field measured at the temperatures indicated. The M versus H shows a jump at low field superimposed on a linear variation, the former associated with the single-ion magnet and the latter with long range anti-ferromagnetism.

To get an insight into the transitions observed in magnetization measurements, we performed specific heat measurements on the small, single-crystal, of $[(CH_3)_2NH_2][Co(HCOO)_3]$as a function of temperature, in various magnetic fields. The as -measured specific heat data, displayed in Fig.4, shows a sharp lambda like transition at 15 K, corresponding with the magnetic transition (cf. Fig. 2), as also has been reported for the Mn system [7]. It is evident from the figure that the specific heat peak that occurs at 15 K, shifts to lower temperature and the area under the specific heat curve reduces with application of magnetic field. To extract the enthalpy change across the magnetic transition, from the specific heat data, a polynomial background function is first subtracted from the specific heat versus temperature plot and peak area evaluated. The variation of enthalpy change obtained at the magnetic phase transition, is plotted as a function of applied magnetic field in Fig.5. Given the two phase magnetic structure of the sample below 15 K, as inferred from the isothermal magnetization studies indicated above, the observed decrease in the enthalpy change with the applied field reflects the decrease in the volume fraction of the antiferromgnetic phase and the corresponding build up of single-ion magnetic structure. The specific heat anomaly at the magnetic transition, similar to results shown in Fig. 4, was earlier observed in a mosaic of single crystals of $(CH_3)_2NH_2][Mn(HCOO)_3]$ using adiabatic calorimetry, as a function of magnetic field [7], and this was attributed to the suppression of the antiferromagnetic transition with the applied magnetic field. We note that while there is a marginal decrease in the antiferromagnetic transition temperature with the applied field, there is a significant change in the enthalpy across the transition that suggests that the explanation in terms of the changing volume fraction of two phase magnetic structure is more tenable.

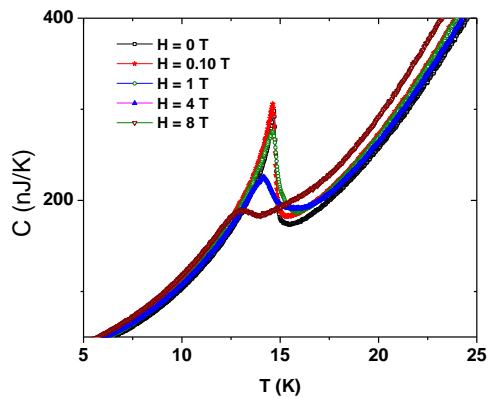

Fig.4. Heat capacity measurements on a single crystal of $[(CH_3)_2NH_2]Co(HCOO)_3$ at various applied magnetic fields as indicated. The peak in heat capacity corresponds to the magnetic transition seen in Fig. 2. A suppression of the peak temperature and broadening of the transition is evident with increase in the strength of the external magnetic field.

The results of specific heat measurements across the order- disorder transition at ~ 160 K, after subtracting the background contribution, is shown in Fig. 6. It is seen that there is considerable hysteresis, pointing to the first order transition, and that the transition temperature is insensitive to the applied field. Similar specific heat anomaly has been seen in Mn [7] Zn and Fe formamide [20] and has been attributed structural transition, associated with the orientational ordering of N, In Fig.7, we plot the ZFC magnetisation data measured in the SQUID- VSM in high temperature range, for measuring fields of 1 T and 4 T. It is seen that there is a distinct jump in the magnetization at 157 K, the temperature that corresponds to the orientational ordering structural transition, riding on the Curie-Weiss behavior. This jump in magnetization, seen for the first time in this family of formamides, and not reported for the Mn, Fe and Ni systems [20], may be attributed to spin state transition of the Co$^{2+}$ ion, associated with the structural transition. Since the jump in magnetization was seen only in the high field experiments at 1 and 4 T, a precise estimation of the change in the Curie constant across the transition was not possible, However, a preliminary analysis from the Curie-Weiss fits to the data measured in low external magnetic field, suggests that effective paramagnetic moment is larger at higher temperatures, implying that the transition of the Co$^{2+}$ state occurs from a lower spin to a higher spin state. Although temperature -induced spin state transition is well known in in Co-oxides [22], it is interesting that in the present system, the abrupt change in the moment at 157 K is coincident with the structural transition. A change in the bond length of Co-O, that is hydrogen bonded to the amino group, across the orientation ordering transition may lead to a change in the spin state of Co$^{2+}$. It would be instructive to look at the bond length changes using infrared and Raman spectroscopies.

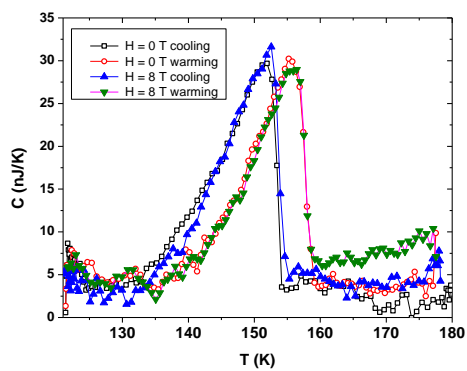

Fig.6. Variation in heat capacity across the order-disorder transition in [(CH$_3$)$_2$NH$_2$]Co(HCOO)$_3$ crystal. The large hysteresis between the heating and cooling cycles is clearly seen. The applied magnetic field ( 8 T ) has no effect on the transition temperature.

## Conclusions

Temperature dependent magnetization and hysteresis measurements in the MOF crystal (CH$_3$)$_2$NH$_2$][Co(HCOO)$_3$], suggest the existence of a two phase magnetic structure within the canted antiferromagnetic phase – as has been reported in Fe system [11]. Evidence for the presence of two co-existing phases is bolstered by the results of specific heat measurements across the magnetic transition at 20 K, in the presence of external magnetic field. In the paramagnetic phase across the structural transition at ~ 160 K, a jump in the magnetization has been seen that has been attributed to spin state transition of Co$^{2+}$.

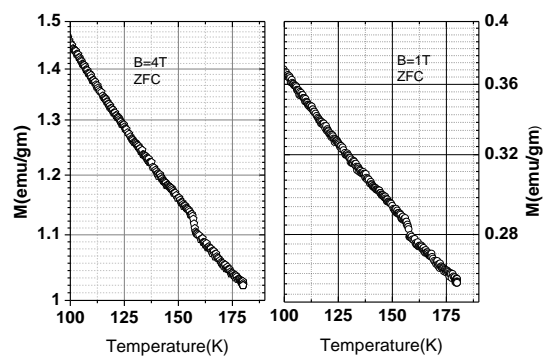

Fig.7 Magnetization versus temperature measured at 4T and 1T in the ZFC protocol in the tiny single crystal at the order-disorder transition, showing the unmistakable jump

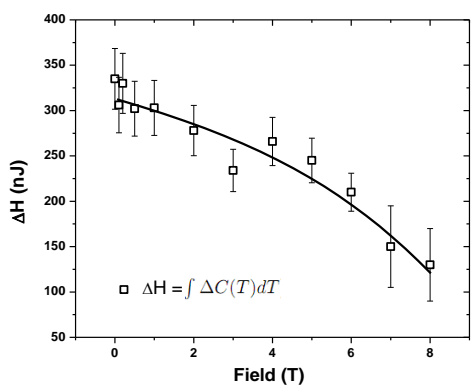

Fig.5. Decrease in the enthalpy change across the antiferromagnetic magnetic transition, with the applied magnetic field. The enthalpy change is obtained by integrating the area under the heat capacity curve, shown in Fig.4, after a polynomial fitted background subtraction.

## Acknowledgements

The authors gratefully acknowledge Dr. A.T. Satya and Dr. A.K Sinha for XRD measurements at the BL-12 beamline of the Indus-II synchrotron. The authors thank Dr. G. Amarendra for making available the SQUID-VSM at the UGC-DAE-CSR node at Kalpakkam for the magnetization measurements.

## Notes and references

[a]Materials Science Group, IGCAR, Kalpakkam, INDIA 603102
[b]Department of Chemistry, Indian Institute of Technology, Kanpur, INDIA 208016